# A Capture-gated Fast Neutron Detection Method

Yi Liu, Yigang Yang, Yang Tai, and Zhi Zhang

*Abstract*—To address the problem of the shortage of neutron detectors used in radiation portal monitors (RPMs), caused by the $^3$He supply crisis, research on a cadmium-based capture-gated fast neutron detector is presented in this paper. The detector is composed of many 1 cm $\times$ 1 cm $\times$ 20 cm plastic scintillator cuboids covered by 0.1 mm thick film of cadmium. The detector uses cadmium to absorb thermal neutrons and produce capture gamma-rays to indicate the detection of neutrons, and uses plastic scintillator to moderate neutrons and register gamma-rays. This design removes the volume competing relationship in traditional $^3$He counter-based fast neutron detectors, which hinders enhancement of the neutron detection efficiency. Detection efficiency of 21.66 $\pm$ 1.22% has been achieved with a 40.4 cm $\times$ 40.4 cm $\times$ 20 cm overall detector volume. This detector can measure both neutrons and gamma-rays simultaneously. A small detector (20.2 cm $\times$ 20.2 cm $\times$ 20 cm) demonstrated a 3.3 % false alarm rate for a $^{252}$Cf source with a neutron yield of 1841 n/s from 50 cm away within 15 seconds measurement time. It also demonstrated a very low (< 0.06%) false alarm rate for a 3.21 $\times 10^5$ Bq $^{137}$Cs source. This detector offers a potential single-detector replacement for both neutron and the gamma-ray detectors in RPM systems.

## [1] INTRODUCTION

Alternative neutron detectors that may replace $^3$He counters, which are popularly used in neutron scattering and homeland security, are being widely researched because of the global supply crisis of $^3$He[1-3]. A new type of composite detector, which can simultaneously measure both neutrons and gamma-rays, is presented in this paper to fulfill the radiation measurement demands in homeland security. Because of the modest cross section of $^3$He in the MeV region [4], moderation is needed to decelerate fast neutrons to slow neutrons. Hence "$^3$He counter + moderator" is the most common mode to measure fast neutrons in RPMs. The volume of the "$^3$He counter + moderator" detector is shared by the volume of the moderator and the volume of the $^3$He counter. The detection efficiency, $\varepsilon_n$, of fast neutrons is roughly determined by the product of two probabilities, $P_m$ and $P_a$, with $P_m$ the probability of decelerating fast neutrons to slow neutrons and $P_a$ the probability of absorbing the decelerated neutrons by $^3$He nuclei.

$$\varepsilon_n = P_m \cdot P_a \qquad (1)$$

It is clear that $P_m$ or $P_a$ will be augmented with increased moderator volume or $^3$He volume. The maximum detection efficiency of the "$^3$He counter + moderator" detector is limited by the compromise between $P_m$ and $P_a$. Fig. 1 gives the simulation results of the detection efficiency curves for $^{252}$Cf neutrons when the volume partition between polyethylene moderator and $^3$He counter is varied. It can be seen that the maximum detection efficiency with a detector size of 9000 cm$^3$ is only 11%, even if the $^3$He pressure is 8 atm.

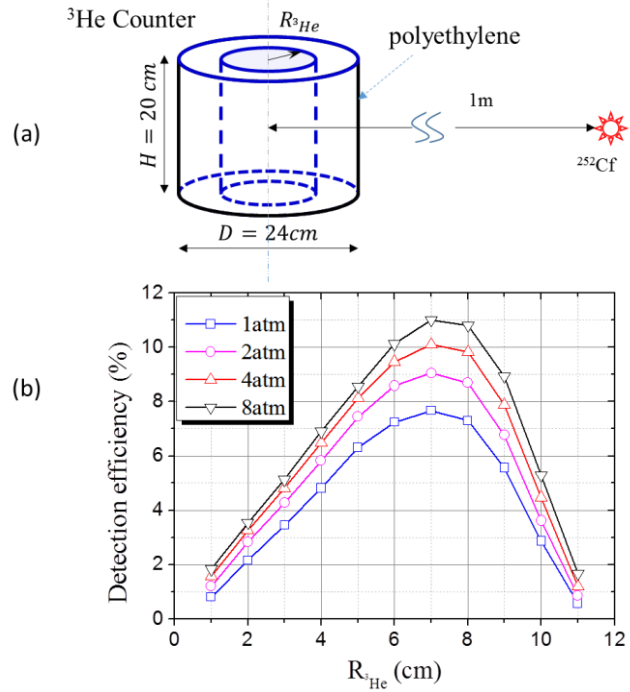

Fig. 1 Simulated intrinsic detection efficiency of the $^3$He detector for $^{252}$Cf neutrons. (a) Structure of a $^3$He detector surrounded by moderator; (b) simulated detection efficiencies for $^{252}$Cf neutrons with different volume partition (varied by the radii of $^3$He counter) between the $^3$He counter and the moderator

The detector researched in this paper uses a plastic scintillator surrounded by $^{113}$Cd foil to replace the "$^3$He counter + moderator". The role of neutron moderation is played by the plastic scintillator, because of its high hydrogen concentration. The role of neutron absorption by $^3$He is played by $^{113}$Cd, which has also a large neutron absorption cross section (2.1 × 10$^4$ barn @ 25.3 meV). The role of signal formation (including ionization and multiplication) played by the $^3$He gas (and other gas mixtures) now is played by the plastic scintillator (and photomultiplier). There are three differences between the "$^{113}$Cd + plastic scintillator" detector and the "$^3$He counter + moderator" detector. First, in the cadmium mode, the volume competing relationship between the moderator and the neutron absorber can be removed, so the maximum detection efficiency will be improved. Second, cascade gamma-rays of MeV energy and internal conversion electrons, as opposed to daughter products of the reaction of $^3$He(n,p)$^3$H, are produced when $^{113}$Cd absorbs slow neutrons. The internal conversion electrons are not further considered because of their poor penetrating capability in cadmium at low energy, so MeV gamma-rays are used to indicate the detection of a neutron. Unlike charged particles, photons always have a chance to escape, so (1) should be modified as:

$$\varepsilon_n = P_m \cdot P_a \cdot P_g \qquad (2)$$

where $P_g$ is the probability that cascade gamma-rays form at least one signal in the scintillator. The ability to perform capture-gated [8-12] detection, which utilizes the true coincidence of the recoil signal and the capture signal from the same neutron to confirm the neutron detection, is the third difference.

[2] DETECTION PRINCIPLE

*A. The structure of the detector*

The detector has a rectangular volume with side length of $S$ and height of $H$. Fig. 2(a) shows the structure of a capture-gated detector, composed of $J$ identical small cells. Fig. 2(b) presents the details of a cell. There are two components for each cell. One component is the plastic scintillator cuboid that plays the roles of moderator and gamma-ray detector. The other is the neutron absorption layer that captures neutrons and produces the secondary photons. $W$ is the side length of the plastic scintillator cuboid, surrounded by the neutron absorption layer of thickness $T$.

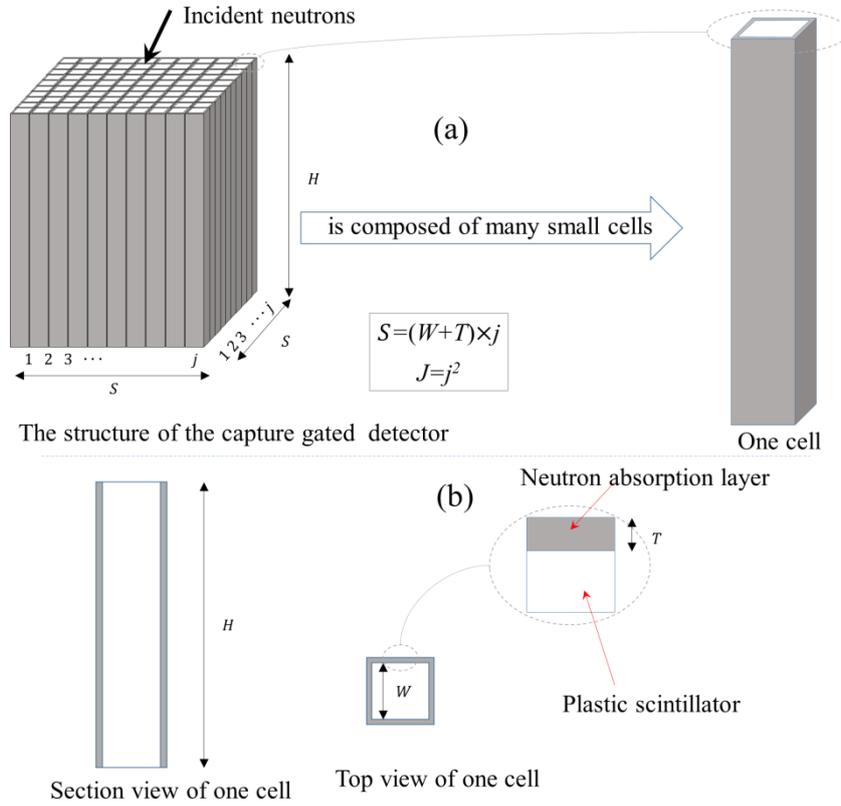

Fig. 2 The structure of the detector. (a) The detector is composed of many small identical cells. (b) The detail of each cell.

*B. The detection of neutrons*

When an incident neutron enters the detector shown in Fig. 2(a), it will undergo three different steps before it might be captured by the detector.

(1) The production of recoil protons: a fast neutron, emitted after the spontaneous fission or (α, n) reactions, has energy in the MeV range and mainly interacts with the detector through (n,n) or (n,n') reactions. Because of the high hydrogen concentration in the plastic scintillator (H : C = 1 : 1.104 for BC408 of Saint-Gobain), most of the energy of the incident neutron gets lost through collisions between the neutron and protons. On average, 11.4 collisions with protons occur before the neutron is absorbed (based on MCNP5 simulation). Due to the fast speed and small free path of the neutron, the time interval between two neighboring collisions is quite short. As shown in Fig. 3, the average time intervals are in the range of several to tens of nanoseconds. Though the plastic scintillator is a fast detector with a short scintillation decay time, when the shaping time of the circuit is hundreds of nanoseconds or greater, recoil protons cannot be discriminated and their scintillation photons would be deemed as photons from the ionization process of one charged particle. Recoil carbon nuclei are also produced. Considering that a $^{12}$C nucleus has 6 times more charge than a proton, its quenching effect will be more serious [13], and hence the equivalent energy deposition will be quite small. If we further notice that the mass of a $^{12}$C nucleus is 12 times heavier than the mass of a proton, the average recoil energy of a $^{12}$C nucleus is about 0.284 (48/169) times than that of a proton's, so the effect of recoil $^{12}$C nuclei is omitted in this discussion.

(2) Neutron capture: when decelerated inside the detector, the neutron is scattered. Because the typical size of $W$ in Fig. 2(b) is about 1 cm ($T$ is even smaller) and $H$ is 10 to 40 cm, then the scattered neutron will cross the neutron absorption layer (see Fig. 2(b)) and move from one cell to the neighboring cell. If the energy of the neutron is not very high, the scattering between the neutron and the proton could be deemed as *S*-wave scattering [13], and the direction of the outgoing neutron will be isotropic in the center-of-mass system. So the travel of the scattered neutron among the cells is not a "one-way" trip, but to some extent looks like a "back and forth" trip. Hence the neutron could be held inside the detector and the number of neutron absorption layers crossed could be large (9 times in simulation if the neutron absorption layer is 0.1 mm thick $^{nat}$Cd with $W$ = 1 cm, $H$ = 20 cm and $N$ = 400). The radius of the "footprint" of the neutron inside the detector would be about 10 cm. Considering that decreased neutron energy usually leads to larger absorption cross section, the scattering process of the neutron inside the detector, which includes deceleration and many crossings through the absorption layers, implies a large absorption probability. $P_A$ is denoted as the probability for an incident neutron to be captured by neutron absorption layers.

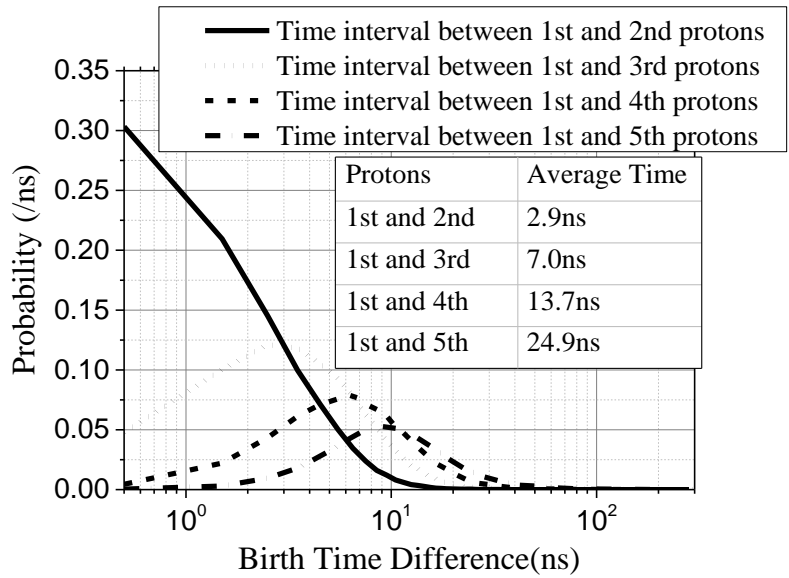

Fig. 3 The distribution of time interval between two neighboring collisions of the neutron with protons

(3) The formation of signal induced by radiation after neutron absorption: the decay of the newly-formed nucleus may emit charged particles (like $^{10}$B(n,α)$^{7}$Li and $^{6}$Li(n,$^{3}$H)α reactions) or gamma-rays and internal conversion electrons (like $^{113}$Cd(n,γ/e)$^{114}$Cd and $^{155,157}$Gd(n,γ/e)$^{156,158}$Gd). If these particles ionize inside the plastic scintillator and form a signal, then the neutron will be detected. $P_S$ is denoted as the probability that emitted particles form a signal inside the plastic scintillator. Because of the short range and serious quenching effect of heavy ions, $P_S$ will be small if $^{10}$B or $^{6}$Li is used as the neutron absorption nuclide. The penetrating capability of MeV photons following neutron capture by $^{113}$Cd or $^{155,157}$Gd is good and their quenching effect is small, so $P_S$ could be large if $^{113}$Cd or $^{155,157}$Gd is used, provided the detection of MeV photons is of high probability. At first glance, $P_g$, the detection probability of MeV photons by the plastic scintillator, which has small atomic number and low density, cannot be large. The fact that $P_S$ is high is because of two reasons: 1) the size of the detector could be large ($H$ = 20 cm and $S$ = 20.2 cm), so $P_g$ is not small (ranging from 29.1% for the 9.04 MeV gamma-ray of cadmium to 71.3% for the 558.5 keV gamma-ray of the most probable emitted photon of cadmium); 2) the excitation energy of the compound nucleus is carried not by one gamma-ray, but several cascade gamma-rays, and hence we can estimate $P_s$ with this equation:

$$P_S = 1-(1-P_g)^{n_g} \qquad (3)$$

where $n_g$ is the number of photons emitted after each capture. For $^{114}$Cd$^*$ formed by the neutron absorption of $^{113}$Cd, $n_g$ is greater than three. Based on the decay scheme of $^{114}$Cd$^*$ [14], the simulation result shows that $P_S$ could be as high as 90%.

C. *The selection of neutron absorption nuclide*

The discussion in Section II.B indicates that the emitted particles after absorption of a neutron should not be heavy ions, and cascade MeV photons are preferred. The candidate nuclide should have:

1.) Large neutron absorption cross section;
2.) High Q value that helps separate the neutron signal from background signals;
3.) Cascade gamma-ray production. The multiplicity of gamma-ray decay of the compound nucleus formed by the neutron absorption improves the gamma-ray detection efficiency, $P_S$, and hence improves the neutron detection efficiency;
4.) Large abundance or could be easily enriched.

Characteristics of several nuclides are shown in Table 1. $^{113}$Cd, $^{155}$Gd and $^{157}$Gd are all good candidates because of their high Q value and the emission of cascade gamma-rays. $^{nat}$Cd was chosen in this research for its convenience to be extruded as a thin foil.

Table 1 Characteristics of neutron absorbing nuclides [15-20]

| Isotope | Cross-section @ 25.3 meV (barn) | Natural abundance (%) | Main reaction products with neutron | Q-value (MeV) |
|---|---|---|---|---|
| $^6$Li | $9.4\times10^2$ | 7.59 | $^3$H, $^4$He | 4.78 MeV |
| $^{10}$B | $3.8\times10^3$ | 19.9 | γ, α, $^7$Li | 2.79 MeV |
| $^{113}$Cd | $2.1\times10^4$ | 12.2 | γ, e$^-$ | 9.04 MeV |
| $^{155}$Gd | $6\times10^4$ | 14.80 | γ, e$^-$ | 8.54 MeV |
| $^{157}$Gd | $2.5\times10^5$ | 15.65 | γ, e$^-$ | 7.94 MeV |

*D. Capture-gated method*

The neutron signal formed in the detector is the result of MeV gamma-ray absorption by the plastic scintillator cuboids. Though the Q-value of $^{113}$Cd(n,γ/e)$^{114}$Cd(9.04 MeV) is far greater than the highest energy of gamma-rays from background radiation (1.46 MeV of $^{40}$K and 2.61 MeV of $^{208}$Tl, for example), the detection of neutrons with this detector cannot be free of background interference. One reason for this is that the response of the plastic scintillator to high energy photons is so poor that the threshold of the discriminator circuit cannot be set too high to lose too many neutron counts. On the other hand, the lowered threshold that helps collect more neutron counts also introduces the interference of background gamma-ray radiation. Another reason is the presence of energetic cosmic rays that cannot be shielded effectively and induces constant count rate. To discriminate neutron signals from background signals, the characteristic behavior of the fast neutron inside the detector is utilized. As shown in Section II.*B*, fast neutrons lose their energy quickly (within 250 nanoseconds) when thermalized. If $t_f$ is the time taken by the neutron from the first proton collision to slow down to the $1/v$ region, and $t_s$ is the time taken by the neutron to be absorbed from when it enters the $1/v$ region (the capture of neutrons before entering the $1/v$ region is omitted because of the low probability), then the time of neutron absorption, $t_a$, from the first proton collision to absorption can be expressed as:

$$t_a = t_f + t_s \qquad (4)$$

To understand the distribution of $t_a$, we observe a neutron in the $1/v$ region for an infinitesimally small time duration, $dt$, from $t$ to $t+dt$. The distance, $dL(t)$, travelled by the neutron within $dt$ is:

$$dL(t) = v(t) \cdot dt \qquad (5)$$

with $v(t)$ the speed of the neutron. Recalling that the structure of the detector, as shown in Fig. 5, is composed of many small identical cells consisting of two different materials, the detector can be approximately deemed as being formed by one "homogeneous" mixture of plastic scintillator and cadmium foil. The linear absorption coefficient of slow neutrons in the detector is then:

$$\mu[v(t)] = \sum_i N_{i_H} \cdot \sigma_{i_H}[v(t)] + \sum_j N_{j_C} \cdot \sigma_{j_C}[v(t)] + \sum_k N_{k_{Cd}} \cdot \sigma_{k_{Cd}}[v(t)] \qquad (6)$$

with $v(t)$ the neutron speed, $N$ the number density, and $\sigma$ the absorption cross section of each nuclide. $i$, $j$, $k$ denote the isotopes of H, C and Cd respectively. $^{113}$Cd contributes the largest portion in (6), which could be further simplified as:

$$\mu[v(t)] \approx N_{^{113}Cd} \cdot \sigma_{^{113}Cd}[v(t)] \qquad (7)$$

The movement of the slow neutron within $dt$ leads to an absorption probability $dP(t)$:

$$dP(t) = \mu[v(t)] \cdot dL(t) \qquad (8)$$

Combining (5) and (7), we get

$$dP(t) = N_{^{113}Cd} \cdot \sigma_{^{113}Cd}[v(t)] \cdot v(t) \cdot dt \qquad (9)$$

Recalling the neutron is in the $1/v$ region,

$$\sigma[v(t)] \cdot v(t) = \sigma_0 \cdot v_0 = \text{constant} \qquad (10)$$

where $\sigma_0$ is the absorption cross section at 25.3 meV and $v_0$ is 2200 m/s, then the probability density of neutron absorption at time $t$ is:

$$\frac{dP(t)}{dt} = N_{^{113}Cd} \cdot \sigma_{^{113}Cd0} \cdot v_0 \quad (11)$$

It is easy to see that d$P(t)$/d$t$ is a constant determined only by the number density of $^{113}$Cd, $N_{^{113}Cd}$, which can be adjusted by modifying $W$ and $T$ in Fig. 2(b). The constant probability density of neutron absorption then leads to the exponential distribution of $t_s$, with the expectation value of $\tau$:

$$\tau = \frac{1}{N_{^{113}Cd} \cdot \sigma_{^{113}Cd0} \cdot v_0} \quad (12)$$

With the typical design, in which $S = L = 20$ cm, $H = 1$ cm, $T = 0.1$ mm, $\tau$ is 2 μs, which is far larger than $t_f$. So (4) can be approximately rewritten as:

$$t_a \approx t_s \quad (13)$$

The above discussion draws the conclusion that the time interval, $t_a$, between the proton signal and the capture signal of the same neutron obeys the following exponential distribution

$$f(t_a) = \tau^{-1} \exp(-t_a / \tau) \quad (14)$$

The short time delay of the capture signal after the proton signal then allows use of the time coincidence technique to confirm the neutron signal and discriminate gamma-ray and cosmic-ray signals. Because each capture signal must be preceded by a proton signal (if no scattering in the environment is considered and capture in the fast region is not considered), the capture-gated method keeps the same detection efficiency as if only the capture signal is used.

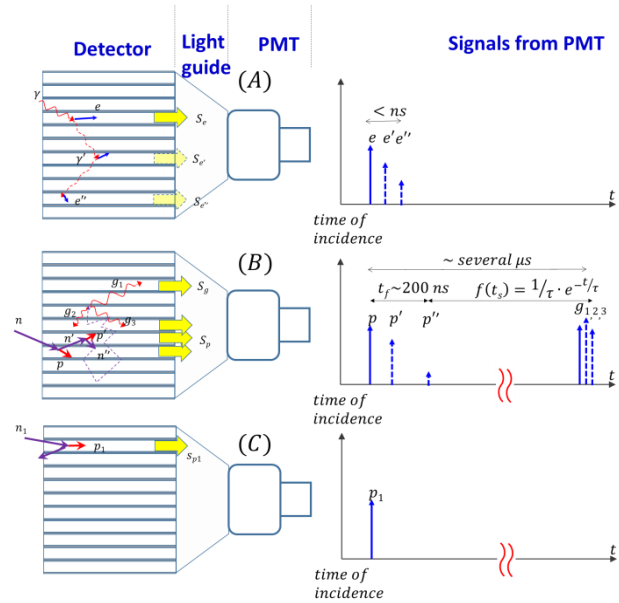

Fig. 4 the time relationship of detector signals induced by photons and neutrons. (A) The photon; (B) The neutron that is captured; (C) The neutron that is not captured.

Fig. 4 shows the time relationship of signals induced by photons and neutrons. In Fig. 4(A), an incident gamma-ray photon is detected and produces one or more secondary electrons (e, e', e''…). Because of the fast speed of the photon, the time difference between the signals of secondary electrons is less than one nanosecond, and hence the signals can be deemed as one signal. In Fig. 4(B), however, an incident neutron, n, first produces several recoil protons. Protons signals, p, p' and p'', are observed to be one signal for the short duration (~ 250 nanoseconds). With the deceleration process of the neutron inside the detector, the neutron is moderated and crosses the neutron absorption layer many times. This neutron is finally (mainly) absorbed by a $^{113}$Cd nucleus of the layer and emits several cascade gamma-rays ($g_1$, $g_2$, $g_3$,…) which produce the signals $g_{1,2,3}$. Because of the fast transition time of gamma-decay, the emission of gamma-rays is simultaneous and their signals are also deemed as one signal. So, for the neutron that is finally captured, there are two signals produced, the proton signal, and the capture signal, which is several microseconds behind the proton signal. Fig. 4(C) shows the neutron signal $n_1$, indicating that the proton signal is not always followed by a capture signal. In this paper, the capture-gated method confirms the detection of the neutron, n, in Fig. 4(B), with the coincidence between the proton signal and the capture signal.

## [3] THE IMPLEMENTATION OF CAPTURE-GATED METHOD

Fig. 5 shows the electronics designed to extract the coupled proton and capture signals. A photomultiplier tube (Hamamatsu CR165) collected the scintillation photons of the detector and converted them to the voltage signals with a custom designed preamplifier using a shaping time less than 1 μs. A custom designed amplifier was used for the further amplification and shaping the signals to narrower than 250 nanoseconds. Output signals were analyzed by a TSCA module (ORTEC, 551) with a user-modified threshold. Signals output by the TSCA were fed to a TAC module (ORTEC 566), with start signals delayed 25 nanoseconds relative to the stop signals. A MCA (ORTEC 919E) was used to convert the amplitude of TAC signals, which reveal the information of time intervals of neighboring signals in the detector, to the time spectrum that will be further analyzed to extract the neutron information.

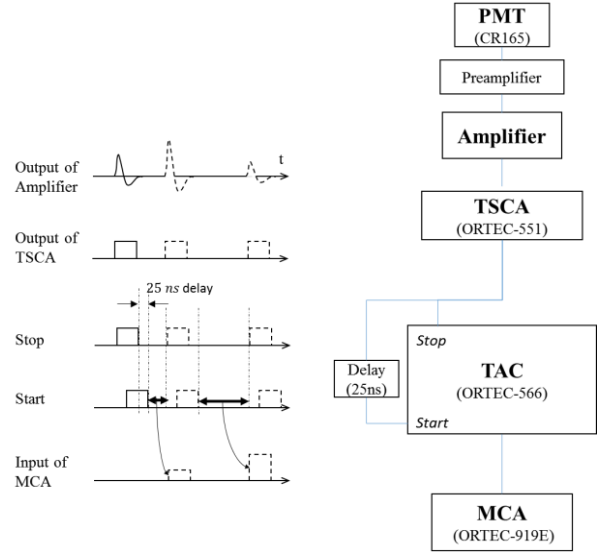

Fig. 5 Electronics of the experiments

In the discussion of Section II, neutron detection is confirmed by the true coincidence of the proton signal and the neutron capture signal. However, proton signals cannot be identified and separated from other signals. In other words, the start signals fed to the "start" of the TAC in Fig. 5 include not only proton signals, but also capture signals and even background signals. In fact, the spectra measured by the MCA were not consistent with the exponential decay shown in (14), which is determined by the true coincidence of the proton signal and the capture signal. The random coincidences of proton, capture and background signals also contribute and make the real time distribution deviate from (14). The formula of the distribution, $f(t)$, of the adjacent signals interval is derived as:

$$f(t) = \frac{1}{1+p} \cdot \exp\left[-mt + mp\tau(e^{-t/\tau} - 1)\right] \times \left\{\left(2m + \frac{1}{\tau}\right) \cdot p \cdot \exp(-\frac{t}{\tau}) + mp^2 \exp(-\frac{2t}{\tau}) + m\right\} \quad (15)$$

where $f(t)$ is the probability density function; $t$ is the time difference of adjacent signals; $m$ is the count rate of non-capture signals (i.e., proton and background signals); $m \times p$, the count rate of capture signals ($p$ is the relative probability of capture signals versus non-capture signals), is also the count rate of the capture-gated proton signals (i.e., the neutron count rate).

## [4] EXPERIMENTAL RESULTS

### A. Layout of the experimental setup

A detector composed of 400 cells, with a sensitive volume of 20.2 cm × 20.2 cm × 20 cm, was irradiated by two different radiation sources, a $^{252}$Cf neutron source with neutron yield of 1841 n/s and a $^{137}$Cs gamma-ray source with activity of $3.21 \times 10^5$ Bq, to investigate the neutron and gamma-ray detection performance of the detector. The distances from the two sources to the detector, $d_n$ and $d_g$, can be changed to simulate different intensities of sources respectively.

### B. Time Interval Distribution and Analysis Method

Fig. 6 presents three time spectra measured with the detector irradiated by background radiation, 662 keV photons of

$^{137}$Cs ($d_g$ = 16 cm) and fission neutrons (also with prompt and delayed photons) of $^{252}$Cf ($d_n$ = 11 cm) respectively. Both the background and $^{137}$Cs spectra show single exponential decay. However, the spectrum of $^{252}$Cf is composed of two parts: fast decay in the "early region" and slow decay in the "later region". The "later region" is the random coincidence contributed by background signals and combined with the $^{252}$Cf signals (so the "later region" of $^{252}$Cf is a little higher than the "later region" of background). The "early region" of $^{252}$Cf is influenced by both the random coincidence of all signals and the true coincidence between proton and capture signals, which accounts for the fast decay of the spectra in the "early region". Because of the coupling between random and true coincidence, the decay in the "early region" is damped. So the measured decay constant, 4.8 μs, which is the exponentially fitted result of the "early region" of the $^{252}$Cf spectrum in Fig. 6, is a little larger than the actual decay constant, 3.7 μs, which is the result fitted with (15).

There are two methods to analyze the spectra in Fig. 6. The first method is fitting them with (15). This method can give the results of the neutron count rate $m_n$, which equals $mp$, the decay time $\tau$, and the total count rate $m(1+p)$ simultaneously. The second method is just integrating counts of the "early region" and "later region". The latter method is simpler and withstands poor counting statistics that usually make the first method fallible in the scenario of portal monitoring when the radiation source is shielded and moving rapidly. The selection of the "early region" and "later region" is to some extent arbitrary. As the rough rule of thumb, the "early region" is from 0 μs to 20 μs (4-5 times the decay time) to include as many true coincidence signals as possible. The "later region" starts 5 μs after the "early region" for the full decay of true coincidence and lasts 20 μs to the end of the spectra. The summation of the counts in the "later region" is denoted by $R_g$ to indicate the existence of the gamma-ray source. $R_n$, which is contributed by the true coincidence of neutron signals and will be used to indicate the existence of the neutron source, is not the summation of counts in the "early region" where the random coincidence also contributes, but is the difference between the total counts of the "early region" and $R_g$.

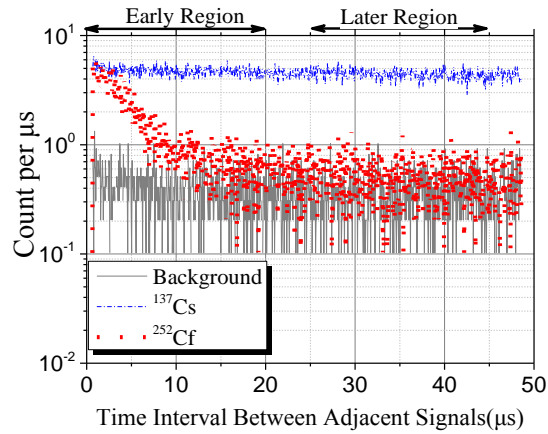

Fig. 6 Time interval distributions when the detector is irradiated with different sources.

*C. Intrinsic Detection Efficiency*

To investigate the neutron detection efficiency, the distance between the neutron source and the detector, $d_n$, is changed to vary the neutron intensity. The relationship between the count rate of neutrons and the solid angle subtended by the detector at the source is:

$$m_n = m_{n,b} + Y_n \cdot \varepsilon_n \cdot \Omega_n \quad (16)$$

where $m_n$ is the neutron count rate fitted with (15), $m_{n,b}$ is the fitted neutron count rate of background, $Y_n$ is the neutron yield of the neutron source, 1841 n/s, and $\Omega_n$ is the solid angle subtended by the neutron detector at the neutron source. $\varepsilon_n$ is the intrinsic detection efficiency of the detector to $^{252}$Cf neutrons. The result shown in Fig. 7 demonstrates very good linearity between the measured neutron count rates versus the varied solid angles. The slope coefficient of the line is 221 ± 1.09 cps/steradian. So the estimated detection efficiency, $\varepsilon_n$, of the detector is 12.00 ± 0.06 %. The equivalent equation of (16) for gamma-rays is:

$$m_g = m_{g,b} + Y_g \cdot \varepsilon_g \cdot \Omega_g \quad (17)$$

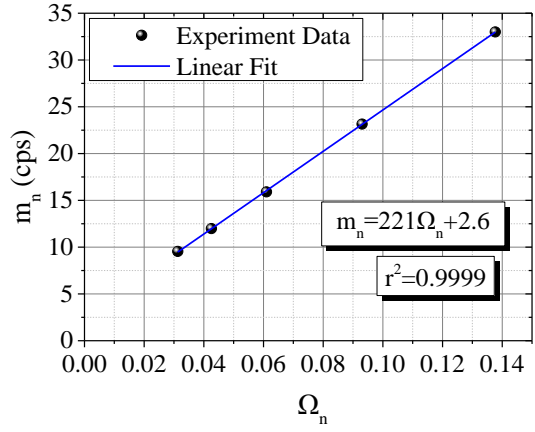

Fig. 7 Measured neutron count rates of the $^{252}$Cf (1841 n/s) source with different solid angles

where $m_{g,b}$ is the gamma-ray count rate of the background; $Y_g$ is the photon yield, $2.73 \times 10^5$ /s, of the $^{137}$Cs source with activity of $3.21 \times 10^5$ Bq; $\Omega_g$ is the solid angle subtended by the neutron detector at the gamma-ray source; and $\varepsilon_g$ is the intrinsic detection efficiency of the detector to 662 keV photons. The results shown in Fig. 8 indicate $\varepsilon_g$ is $11.6 \pm 1.4$ %.

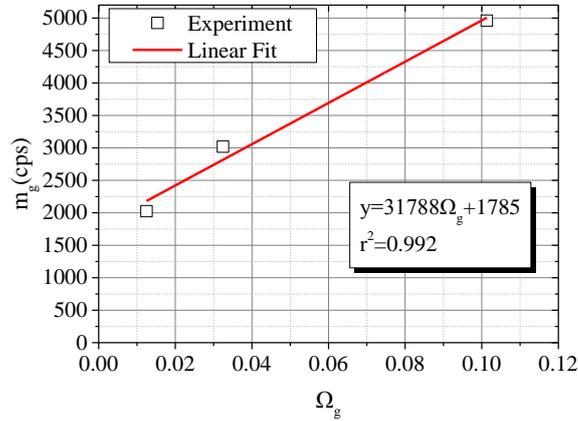

Fig. 8 Measured photon count rates of the $^{137}$Cs source ($3.21 \times 10^5$ Bq) with different solid angles

*D. Simultaneous alarming of neutron and gamma-ray sources*

The discussion in Section IV.B indicates that $R_n$ and $R_g$ could be used to indicate the potential existence of neutron and gamma-ray sources. To evaluate the performance of this detector, the time interval distributions of background, neutron source at different distances, and gamma-ray source at different distances were measured for a total time of 1800 seconds respectively. For each measurement, the time distribution is recorded as 1800 spectra of 1 second measurement duration. The values of $R_n$ and $R_g$ of the 1800 spectra were calculated and tallied to form the count distributions of the neutron and the gamma-ray sources, as shown in Fig. 9 and Fig. 10 respectively, to find the false alarm rates.

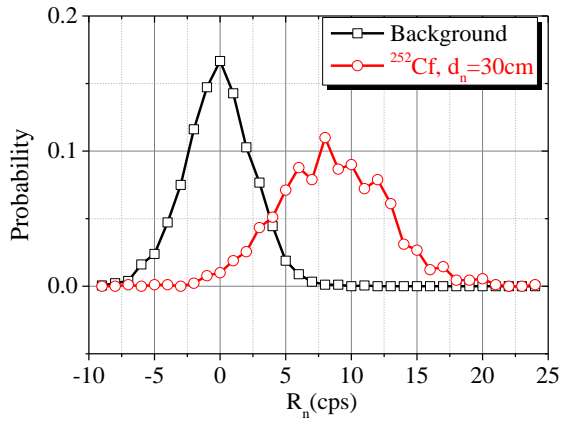

Fig. 9 Count distributions of $R_n$ of background and $^{252}$Cf neutron source ($d_n$ = 30cm, single measurement time is 1 second)

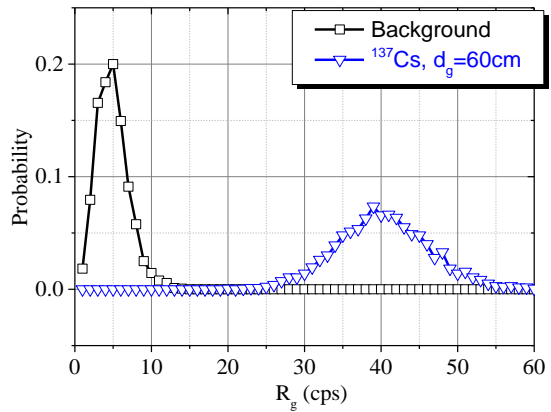

Fig. 10 Count distributions of $R_g$ of the background and the $^{137}$Cs gamma-ray source ($d_g$ = 60cm, single measurement time is 1 second)

Table 2 and Table 3 give the calculated false alarm rates of the neutron source and gamma-ray source for nine different scenarios with varied measurement times. It can be seen that the false alarm rates of the neutron source could be enhanced by prolonged measurement time or reduced $d_n$. There can be a false alarm rate of 19.0 % within 1 second measurement when $d_n$ is 30 cm. The same false alarm rate requires a 10-second measurement time if the gamma-ray source is also present and positioned 40 cm away. The identification of the gamma-ray source is significantly better as the largest false alarm rate is 0.06 % even in the worst case.

Table 2 False Alarm Rates of the Neutron Source (%)

| No. | Distance (cm) | | Measurement Time (s) | | | | | |
|---|---|---|---|---|---|---|---|---|
|  | $d_n$ | $d_g$ | 1 | 5 | 10 | 15 | 20 | 60 |
| 1 | 11 | ∞ | 0.1 | 0 | 0 | 0 | 0 | 0 |
| 2 | 30 | ∞ | 19.0 | 0 | 0 | 0 | 0 | 0 |
| 3 | 50 | ∞ | 54.6 | 20.1 | 8.6 | 3.3 | 0 | 0 |
| 4 | 30 | 40 | 68.4 | 32.8 | 18.9 | 9.2 | 10.0 | 0 |
| 5 | 30 | 60 | 52.6 | 13.9 | 1.7 | 0 | 0 | 0 |

Table 3 False Alarm Rates of the Gamma-ray source (%)

| No. | Distance (cm) | | Measurement Time (s) | | | | | |
|---|---|---|---|---|---|---|---|---|
|  | $d_n$ | $d_g$ | 1 | 5 | 10 | 15 | 20 | 60 |
| 6 | ∞ | 40 | 0 | 0 | 0 | 0 | 0 | 0 |
| 7 | ∞ | 60 | 0 | 0 | 0 | 0 | 0 | 0 |
| 8 | 30 | 40 | 0 | 0 | 0 | 0 | 0 | 0 |
| 9 | 30 | 60 | 0.06 | 0 | 0 | 0 | 0 | 0 |

## [5] CONCLUSION

Research on a fast neutron detector which uses the capture-gated method has been presented in this paper. The detector is constructed with many small cells composed of plastic scintillator cuboids surrounded by thin cadmium layers. The moderator and neutron absorber can be deemed as homogeneously mixed, and hence the drawback of limited detection efficiency in the traditional $^3$He counter based solution is overcome by removing the volume competing relationship between the moderator and the neutron absorber. The neutron detection efficiency is measured to be 12.00 ± 0.06 % with a 20.2 cm × 20.2 cm × 20 cm detector. A larger detector, with the size of 40.4 cm × 40.4 cm × 20 cm, was also constructed and it showed a detection efficiency of 21.66 ± 1.22%. The equivalent count rate of 1 ng $^{252}$Cf neutron source with $d_n$ = 200 cm is expected as 3.45 cps, which is larger than the 2.5 cps requirement, in the RPM system.

This single detector can both measure neutrons and gamma-rays simultaneously. It shows the potential of replacing not only the $^3$He counter, but also the gamma-ray detector in current RPM systems.